\documentclass{article}

\usepackage[english]{babel}

\usepackage[a4paper,top=2cm,bottom=2cm,left=3cm,right=3cm,marginparwidth=1.75cm]{geometry}

\usepackage{amsmath}
\usepackage{graphicx}
\usepackage[colorlinks=true, allcolors=blue]{hyperref}
\usepackage[normalem]{ulem}

\title{Radioactive nuclei for $\beta^+\gamma$ PET and theranostics: \\ selected candidates}
\author{Tomasz Matulewicz \\{Faculty of Physics, University of Warsaw, Poland}}

\begin{document}
\maketitle

\begin{abstract}
PET is an established medical diagnostic imaging method. 
Continuous improvements are aimed at refining image reconstruction, reducing the amount of radioactive tracer and combining with targeted therapy.
TOF-PET provides the localization of the tracer through improved time resolution, nuclear physics may contribute to this goal via selection of radioactive nuclei emitting additional $\gamma$-rays.
This additional radiation, when properly detected, localizes the decay of the tracer at the line of response determined by two detected 511 keV quanta.
Selected candidates are presented.
Some are particularly interesting, as they are strong candidates for theranostic applications.
\end{abstract}

\bigskip 

{\it 
\null\hfill Nowadays, it is obvious that society largely benefits from
\newline 
\null\hfill the large investments done in basic Nuclear Physics  research. \newline
\null\hfill Recent achievements in particle- and radio-therapy  within \newline 
\null\hfill the new paradigm of theranostic approach  are some of   \newline 
\null\hfill the most striking examples of the benefits  \newline  \null\hfill  from Nuclear Physics. \cite{NUPECC}
 }

\section{Introduction}

Positron Emission Tomography (PET) is nowadays a standard medical diagnostic imaging technique. 
The average range of a positron from the $\beta^+$ decay of a radionuclide (tracer) is of the order of mm in the tissue. 
The spatial range of positrons following $\beta^+$ decay in water depends on the emission spectrum and is well described by the sum of two exponents \cite{Craig}.
The compounds with radioactive nuclei $^{18}F$ are predominantly used (above 90\% of performed scans), but other tracers (containing $^{11}C$ or $^{68}Ga$) are found to have superior imaging properties \cite{Sharma2013} in certain cases.
For $^{18}F$ the FWHM of the distribution is 0.102 mm \cite{Craig}; the range is below 2.4 mm \cite{Huang}.
The annihilation of the positron with the electron creates usually two gamma quanta of 511 keV energy, emitted back-to-back, thus conserving null momentum of annihilating system. 
The mean free path of 511 keV photons in water is about 10.4 cm (computed from \cite{NIST}).
Detection of both 511 keV photons in coincidence in scintillation crystals enables the possibility to obtain the Line of Response (LoR) between the geometrical positions of fired detectors. 
The intersection of multiple LoRs provides, in the first approximation, the location of the tracer. 

This standard procedure augmented by additional corrections due to attenuation and scattering of $\gamma$ quanta) does not account for the position of the annihilation along LoR. 
This information is, in principle, stored in the time difference between 511 keV quanta interaction in detectors. 
Time-resolving power of detectors and associated electronic circuits are the key issue in this approach, as 3 cm difference along LoR corresponds to 200 ps. 
Important progress has been observed in this technology, as the time resolution of commercial scanners improved from 1000 ps two decades ago, to 550 ps a decade ago and 220 ps in currently available devices (four-fold increase of the axial dimension should also be noted). 
The Time-of-Flight (TOF) PET scanner, instead of full LoR, employs the concept of segment of response, determined by the time bin.
The improvement of time-resolving power leads to the reduction of the background and allows the possibility to reduce the dose of tracer, important in several cases
\cite{TOF-PET}.
Coincidence resolving time in 10 ps range would lead to enormous improvement, allowing for on-line image reconstruction.
The roadmap to achieve this goal has been recently published \cite{10ps}.

The decay vertex (lying close to the LoR) might be determined from the radioactive ($\gamma$) decay of the daughter nucleus, provided the lifetime is short enough to stay within a coincidence window.
The “third” photon should be detected via Compton scattering following the photoelectric absorption of the scattered $\gamma$ quantum. 
This method provides localization of the decay vertex through the intersection of the reconstructed Compton cone with LoR on an event-by-event basis.
The “third” photon should be around 1 MeV energy to increase the probability for Compton effect in detector material. 
Improved localisation of the vertex in this $\beta^+\gamma$ technique leads to the reduction of the global radioactive dose administered for the diagnosis, although some dose would come from this “third” (and additional) gamma.

\begin{figure}[t]
    \centering
    \includegraphics[width=0.7\textwidth]{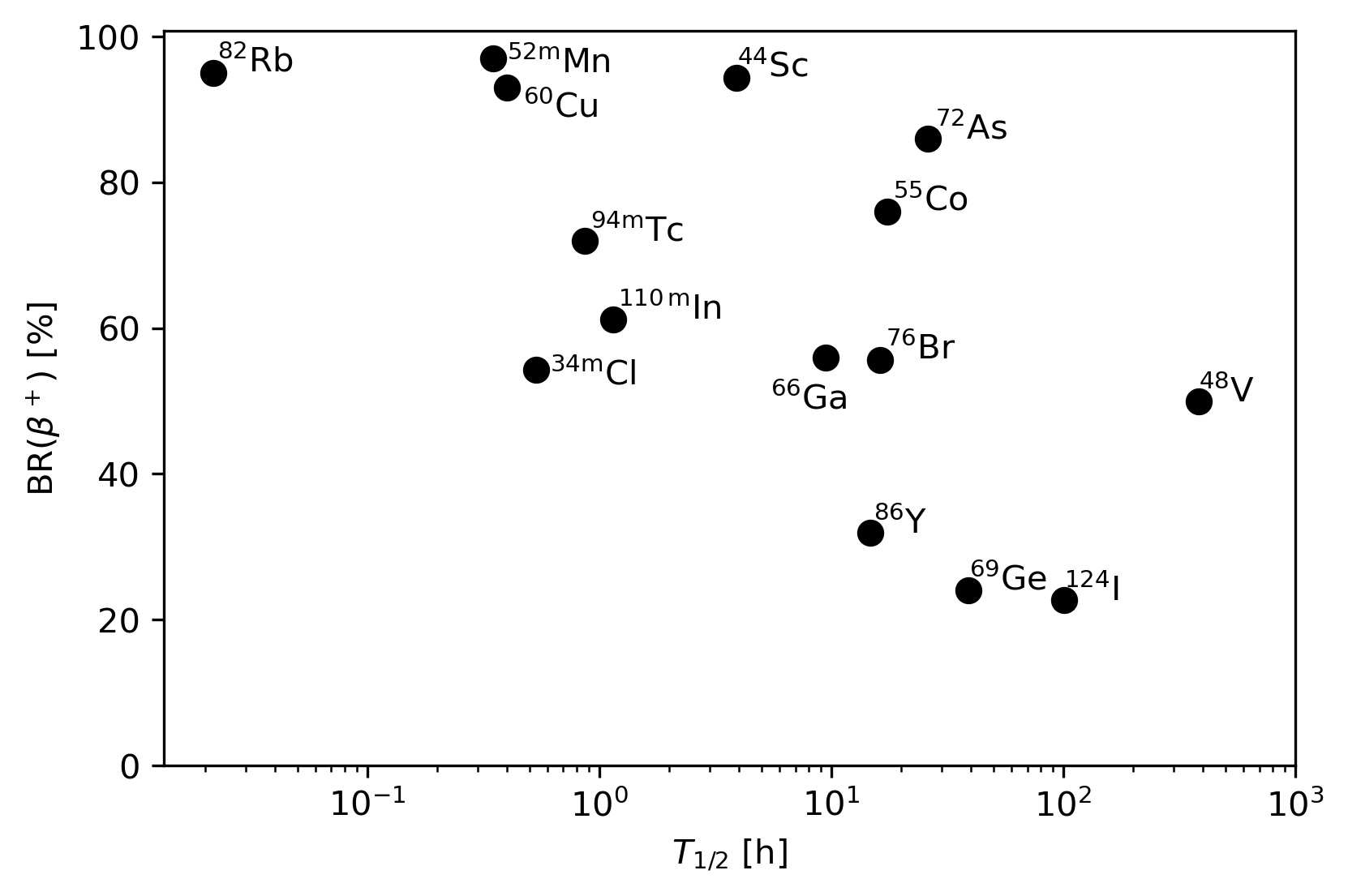}
    \caption{$\beta^+$ branching ratio versus halflife of nuclei - candidates for $\beta^+\gamma$ PET. }
    \label{T12_BR}
\end{figure}

\section{Candidates for radionuclides in  {$\beta^+\gamma$} PET}

The  $\beta^+\gamma$ PET technique has been under development for more than two decades \cite{Sitarz}. 
The approach seems to be more effective compared to the 3$\gamma$ decay of ortho-positronium, as this process is rare (typically below 1{\%}) and requires more sophisticated treatment of the data. 
Suitable candidates for radionuclides to be used in  $\beta^+\gamma$ PET should fulfil several conditions:
\begin{enumerate}
    \item Suitable lifetime (long enough for diagnostic procedure and short enough to reduce unwanted irradiation of the body after the medical procedure),
    \item High $\beta^+$ branching ratio (electron capture as the natural concurring process), 
    \item Energy of the $\gamma$ radiation favouring Compton scattering (i.e. $\simeq1$MeV),
    \item Multiplicity of other $\gamma$-rays should in principle be low.
\end{enumerate}

\begin{figure}[h]
    \centering
    \includegraphics[width=0.65\textwidth]{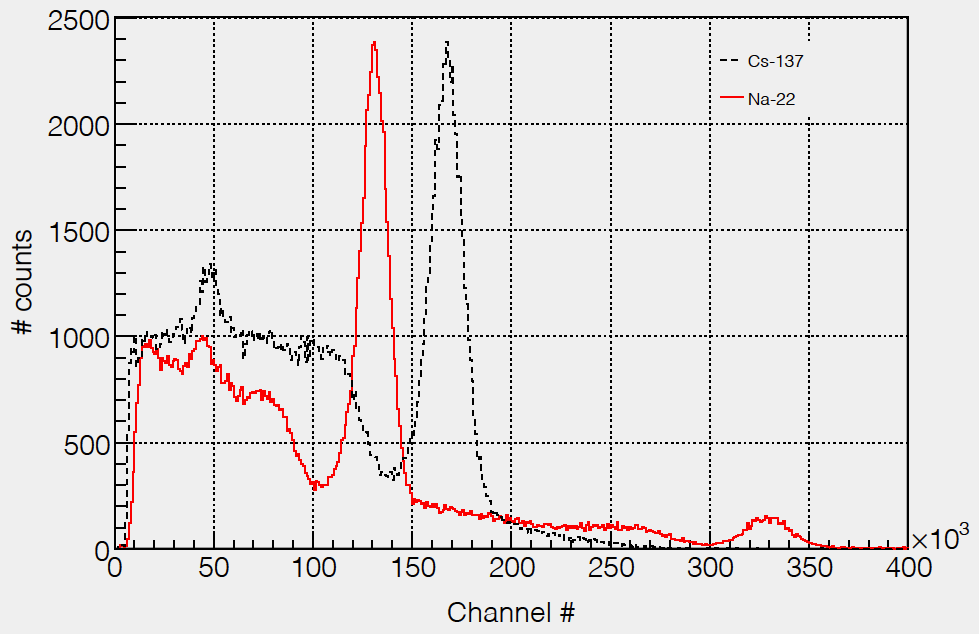}
    \caption{Energy spectra of 5$\times$5$\times$20 mm$^3$ LYSO crystal for $^{22}$Na and $^{137}$Cs radioactive sources. The spectra are normalized at 511 keV and 662 keV peaks (from \cite{PhDBergen}).}
    \label{511_662}
\end{figure}

\begin{figure}[ht]
    \centering
    \includegraphics[width=0.7\textwidth]{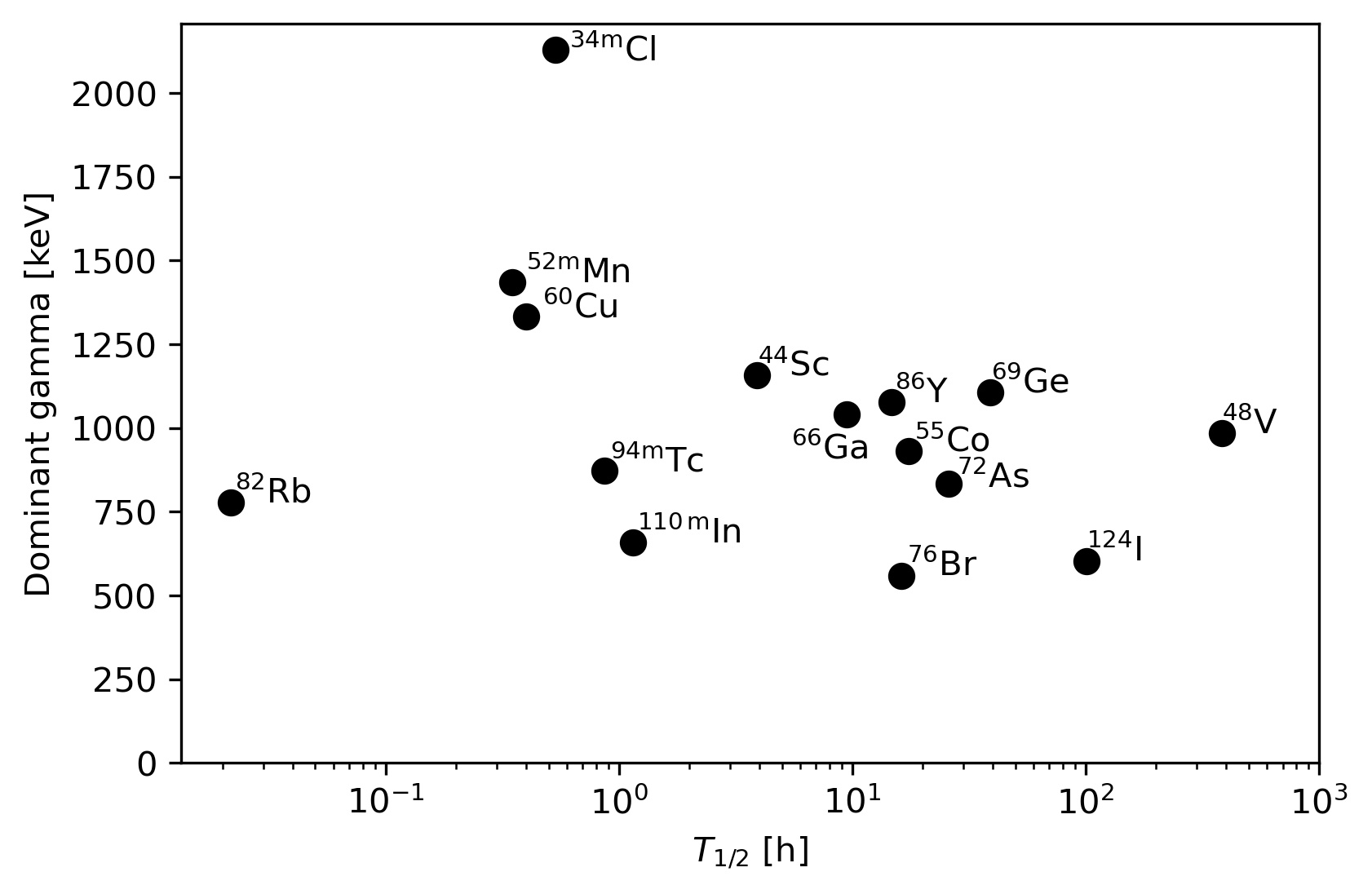}
    \caption{Energy of most probable $\gamma$-ray versus halflife of nuclei - candidates for $\beta^+\gamma$ PET. }
    \label{T12_Egamma}
\end{figure}

The candidates for those radionuclides have been recently reviewed by M. Sitarz et al. \cite{Sitarz}. 
The basic properties of those nuclei are visualized in Figures \ref{T12_BR} and \ref{T12_Egamma}.
Their lifetime is in the order of hours. 
The $\beta^+$ branching ratio of considered nuclei (see Fig.\ref{T12_BR}) makes them prospective candidates. 
Several of them have branching ratios well above 80\%, so almost all radioactive decays could contribute to the $\beta^+$ emission, which is essential to the PET technique.
The energy of the $\gamma$ radiation emitted after the decay from the excited state of the daughter nuclei is an important factor for the envisaged technique. 
The lifetime of the excited state should allow for coincident measurement, so lifetimes in the range of ps are needed.
In fact, this is the case of all considered nuclei.
The $\gamma$-ray should be Compton-scattered, so $\simeq1$MeV $\gamma$-rays are best suited.
The energy spectra (Fig.\ref{511_662}) for $^{22}$Na and $^{137}$Cs calibration sources registered by LYSO detector (typical for PET devices) shows how pronounced is the rise of Compton-scattering events when the $\gamma$-ray energy increases from 511 keV (annihilation of $\beta^+$ after the decay of $^{22}$Na) to 662 keV (after $^{137}$Cs decay).
The energy of the most abundant $\gamma$-ray emitted after the $\beta^+$ decay of considered candidates is in the region of $\sim$1 MeV (see Fig.\ref{T12_Egamma}).

\begin{figure}[ht]
    \centering
    \includegraphics[width=0.7\textwidth]{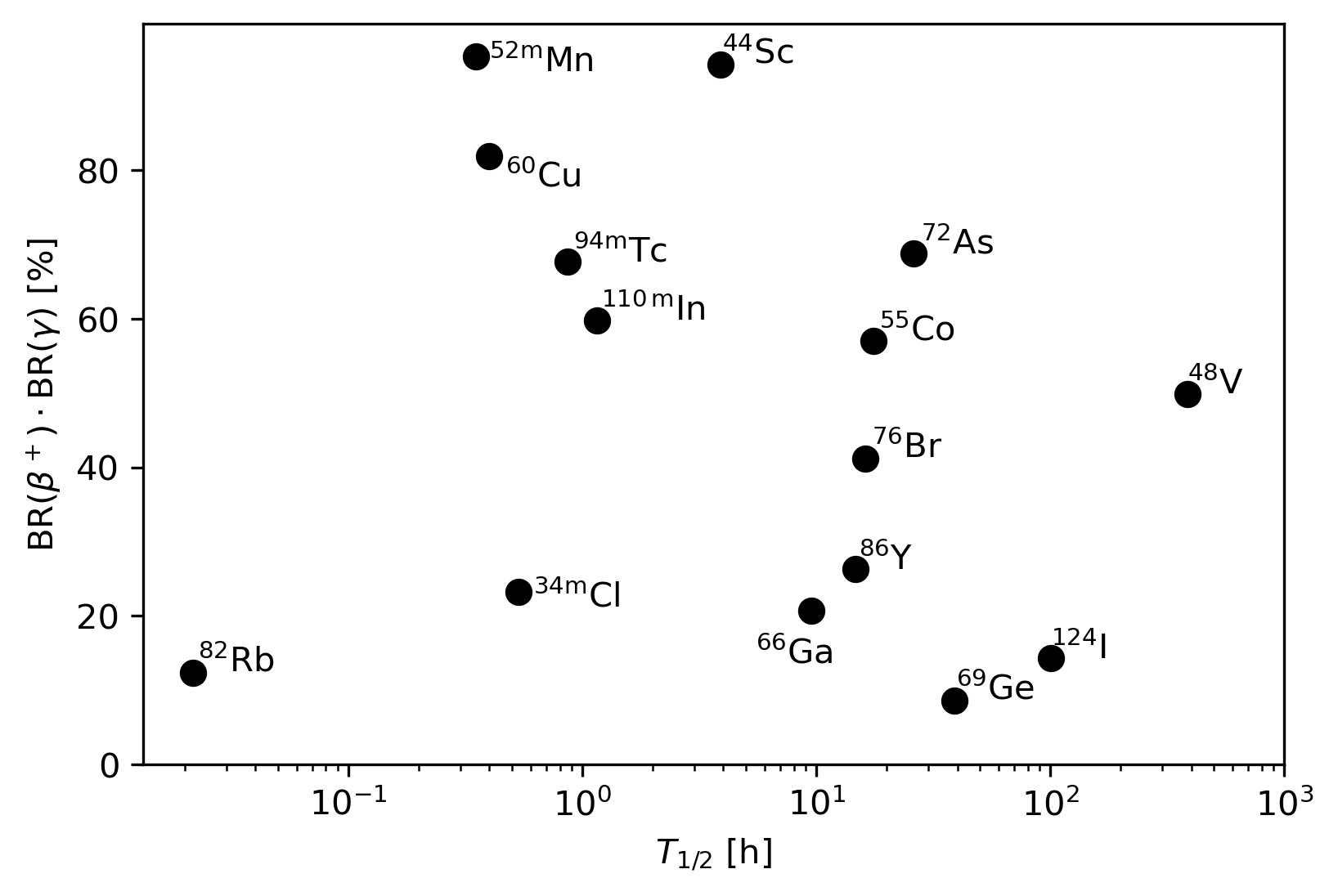}
    \caption{Branching ratio of the most probable $\gamma$-ray multiplied by $\beta^+$ branching ratio versus halflife of nuclei - candidates for $\beta^+\gamma$ PET. }
    \label{T12_2BR}
\end{figure}

The branching ratio of $\beta^+$ decay is the key factor for the standard PET operation method (almost 100\% for$^{18}$F).
For the $\beta^+\gamma$ PET, the product of $\beta^+$ branching ratio and the fraction of so-called "third" $\gamma$ emission in the decay of the daughter nucleus, determines the efficiency of the process with respect to the dose of the tracer. 
Evidently (see Fig.\ref{T12_2BR}), $^{44}$Sc, $^{52m}$Mn and $^{60}$Cu are three most suitable candidates.

The three conditions mentioned above do not allow selection of the best nuclei for this technology.
Apart from making suitable radio-pharmaceutic (chemistry), key issues are other properties of the decay and the method of producing the resulting sample. 
A very important aspect is the purity of the sample - the amount of co-produced (usually unwanted) radionuclides.
For example, production of $^{34m}$Cl requires $\alpha$ beam of $\sim$65 MeV kinetic energy, beyond the range available in typical medical accelerators. Also, the decay of $^{34m}$Cl populates many other excited states decaying via energetic $\gamma$-rays, which complicates the analysis and generates unwanted radiation effects. 
So, in spite of the high energy of emitted "third" $\gamma$-ray and branching ratio above 50\%, the $^{34m}$Cl isotope is not one of the most appropriate candidates.

The production of medically interesting radioisotopes might provide unwanted activities of the same chemical element, especially in the case of non-monoisotopic targets, thus impossible to separate chemically.
Short lifetime of unwanted activity would make the sample clear after a time, but this is rarely the case.
Selective reactions, employing targets enriched with the appropriate isotope, are commonly used and efficient methods that have been developed for re-use of the target material.
The production cross section for different reaction channels have been measured intensively in the past decades and numerical tools have been developed to evaluate the expected yields \cite{RYC}. 
However, not all production routes have been fully experimentally verified or optimized and experimental activity is welcome.
The numerous commercially available accelerators (now around 1500 worldwide \cite{Ncyclotrons}) provide the natural place to supply radioisotopes for this prospective PET technique.
Their maximum energy is usually limited to $\leq$20 MeV protons, so those among the perspective nuclei for $\beta^+\gamma$ PET, which can be produced with this beam (or appropriately scaled in energy deuterons and $\alpha$-particles) are naturally preferred. 
Among them are $^{44}$Sc, $^{48}$V (for slow metabolic processes), $^{55}$Co, $^{60}$Cu, $^{66}$Ga, $^{94m}$Tc  and $^{124}$I (unfortunately low $\beta^+$ branching ratio).
Some others may be obtained via nuclear generators, like $^{82}$Rb supplied from $^{82}$Sr (T$_{1/2}$=25.4 d).
Several of those radionuclides were already used in diagnosis of certain diseases.

\section{Nuclei for theranostic applications}

The theranostic approach aims to combine diagnosis and therapy to provide effective treatment at the very early stage of cancer \cite{Krolicki}. 
The role of Nuclear Physics is to develop specific radioisotopes, providing diagnostics functionality together with therapeutic effect, so this activity has gone beyond standard $^{18}F$ PET tracers.
These requests can be met with single radionuclide for imaging and therapy (like $^{117m}$Sn) or by pairs of isotopes having the same chemical properties (like $^{44}$Sc/$^{47}$Sc, $^{64}$Cu/$^{67}$Cu and others) or similar (like $^{99m}$Tc/$^{188}$Re \cite{TcRe}).
Evidently, $^{44}$Sc seems to be a natural link between $\beta^+\gamma$ PET and theranostic approach.
The $^{44}$Sc/$^{47}$Sc pair is the subject of wide-front research.
The production of these radionuclides is studied intensively by very different methods (reports from only 5 last years):
\begin{enumerate}
    \item Photonuclear reactions \cite{Sc-gamma1, Sc-gamma2},
    \item Neutron irradiation of natural calcium in reactor \cite{Sc-reactor},
    \item Low-energy proton and deuteron beams \cite{Sc-hydrogen1, Sc-hydrogen2, Sc-hydrogen3},
    \item $^3$He reactions at low energies \cite{Sc-helium} and
    \item Spallation reaction induced by $\sim 1$GeV protons \cite{Sc-spallation}.
\end{enumerate}
There is also important progress in chemistry and medical applications of this pair of nuclei.
The production, chemistry and in-vivo studies of Sc-isotopes were reviewed few years ago \cite{Sc-review1, Sc-review2}.
Currently, around 20 papers are published yearly on the subject of production and medical applications of this theranostic pair of scandium nuclei, so any review soon might become outdated.

The progress in the development of other important theranostic pairs is recently significant, in particular concerning radioactive Cu isotopes.
The photonuclear production of $^{67}$Cu was \sout{recently} found to be an efficient route \cite{photo67Cu}.
In this case the final product is free of co-produced isotopes, populated otherwise in proton, deuteron and alpha-induced reactions on zinc targets.
The use of highly-enriched targets limits the production of long-lived $^{64}$Cu, what was recently demonstrated \cite{70Zn-67Cu} in the measurement of deuteron-induced reactions on $^{70}$Zn.
This progress is particularly important, as the therapeutic application of $^{67}$Cu has recently been demonstrated \cite{Medical67Cu} and found efficacious in certain prostate cancers.

\section{Conclusions}

Not only nuclear aspects would select the best radioisotope for {$\beta^+\gamma$} PET, as many other properties (e.g. chemistry) and technical restrictions would play a role.
Special medical requests and imaging properties are also of primary concern.
The pure nuclear aspects favour $^{44}$Sc and $^{60}$Cu, as their production is also possible in commercial PET-supplying cyclotrons.
However, many  others have already been used in practical diagnosis procedures and deserve further studies.
The progress in theranostic application of $^{44}$Sc/$^{47}$Sc pair is intensively studied and different production routes were recently evaluated.
Other dedicated combinations are also of interest, particularly as their therapeutic efficiency was demonstrated.
The demanding detection conditions of {$\beta^+\gamma$} process can be effectively realized in a total-body PET \cite{TotalPET}, where the large geometrical acceptance of the device would be the key factor for making fast measurements with low-dose.

\bigskip{}
{\bf Acknowledgments}

I am grateful to dr. Mateusz Sitarz and the late professor Jerzy Jastrz\c{e}bski for pointing to me their vision of medical applications of nuclear physics. 
The help of Joanna Matulewicz in making graphics is acknowledged.

\end{document}